\documentclass[prl,twocolumn,10pt]{revtex4}

\usepackage{amsmath}
\usepackage{dcolumn}
\usepackage[cp1251]{inputenc}
\usepackage[english]{babel}
\usepackage{epsfig}
\usepackage{graphics}
\usepackage{natbib}
\bibpunct{[}{]}{,}{a}{}{;}

\newlength{\twocolumnwidth}
\providecommand{\preprintmarginvar}[1]
{\hspace{#1\columnwidth}\hspace{-#1\twocolumnwidth}} 
\providecommand{\preprintmargin}{\preprintmarginvar{0.4}} 

\providecommand{\Eq}[1]{Eq.\ (\protect\ref{eq:#1})}
\providecommand{\Eqs}[1]{Eqs.\ (\protect\ref{eq:#1})}

\providecommand{\hide}[1]{} 
\providecommand{\eqlabel}[1]{\addtocounter{equation}{1}
	\tag{\arabic{equation}{\scriptsize \{eq:#1\}}}\label{eq:#1}} 
 
\providecommand{\FM}{^{(-)}}

\providecommand{\etal}{{\em et al.\/}}

\providecommand{\generalbracket}[4]{\protect\ensuremath{#1#2#4#1#3}}

\providecommand{\Qbracket}[2]%
 {\protect\generalbracket{#1}{\langle}{\rangle}{#2}} 
\providecommand{\Vbracket}[2]%
 {\protect\generalbracket{#1}{\langle 0#1|}{|0#1\rangle}{#2}} 
 
\providecommand{\Kbracket}[2]%
 {\protect\generalbracket{#1}{|}{\rangle }{#2}} 
\providecommand{\Bbracket}[2]%
 {\protect\generalbracket{#1}{\langle }{|}{#2}} 
\providecommand{\Ebracket}[3]%
 {\protect{#1\langle #2 #1| #3 #1\rangle}} 
\newlength{\QQlength}
 
\providecommand{\vacavbracket}[2]%
 {#1\langle 0#1|#2#1|0#1\rangle} 
\providecommand{\Z}{ }
\providecommand{\formA}[1]%
 {{\renewcommand{\Z}{&}\begin{aligned}#1\end{aligned}}}
\providecommand{\formG}[1]%
 {{\renewcommand{\Z}{ }\begin{gathered}#1\end{gathered}}}
\providecommand{\eqM}[2]%
 {{\renewcommand{\Z}{ }\begin{multline}\preprintmargin #1\preprintmargin #2\end{multline}}}
\providecommand{\eqMW}[2]%
 {{\renewcommand{\Z}{ }\begin{multline}#1 #2\end{multline}}}
\providecommand{\eqA}[2]%
 {\protect{\begin{align}{{\renewcommand{\Z}{&}
 \begin{aligned}#1\end{aligned}}}#2\end{align}}}
\providecommand{\eqG}[2]%
 {\protect{\begin{gather}{{\renewcommand{\Z}{ }
 \begin{gathered}#1\end{gathered}}}#2\end{gather}}}
 
\renewcommand{\eqlabel}[1]{\label{eq:#1}}

\begin{document}

\title{Efimov states in atom-molecular collisions}

\author{M.A.\ Efremov,$^{1}$ L.\ Plimak,$^1$ B.\ Berg,$^1$
M.Yu.\ Ivanov,$^2$ and W.P.\ Schleich$^1$}

\affiliation{$^1$Institut f\"ur Quantenphysik, Universit\"at Ulm,
89069 Ulm, Germany \\$^2$Steacie Institute for Molecular Sciences,
NRC Canada, 100 Sussex Drive, ON Ottawa, K1A 0R6 Canada}
\email{max.efremov@gmail.com}



\date{\today}

\begin{abstract}

We analyse scattering of a heavy atom off a weakly bound molecule
comprising an identical heavy and a light atom in the
Born-Oppenheimer approximation. We focus on the situation where the
heavy atoms are bosons, which was realized in several experiments.
The elastic and inelastic cross sections for the atom-molecular
scattering exhibit a series of resonances corresponding to
three-body Efimov states. Resonances in elastic collisions are
accessible experimentally through thermalization rates, and thus
constitute an alternative way of observing Efimov states.

\end{abstract}

\maketitle

\noindent{\em Introduction.\/}--- The Efimov effect \cite{Efimov} is
the emergence of a large number of weakly bound three-body states if
at least two of the three two-body subsystems exhibit a weakly bound
state or resonance. This implies that the two-body scattering length
$a_0$ is much larger than the characteristic radius of the two-body
interaction $r_0$. The number of three-body states is proportional
to $\ln(|a_0|/r_0)$. In the resonant limit $|a_0|\rightarrow
\infty$, the energies of the three-body states form a geometric
sequence, with the common ratio determined by the exponent
$\exp(2\pi/s_0)$. The parameter $s_0$ depends on the masses of the
particles and the number of ``participating'' resonant two-body
interactions (two or three) \cite{Efimov,Esry,Hammer}.

First candidates for the Efimov effect were halo nuclei
\cite{Fedorov-Jensen} and the helium trimer \cite{He-experiment}. In
these systems the scattering length is exceedingly large by nature
and fixed. However, it is experimentally beneficial to have control
of the scattering length so as to observe Efimov states emerging
with changing $a_0$. This opportunity is provided by ultracold
atomic gases which are now regarded the most promising candidates.
Tuning the scattering length in an external magnetic field near a
Feshbach resonance was used to observe a Efimov resonance in an
ultracold gas of Caesium atoms for the negative scattering length
\cite{Grimm}. In this experiment, resonant three-body recombination
losses were observed when the strength of the two-body interaction
$a_0$ was varied. The resonance was attributed to a Efimov state.
More recently \cite{Knoop}, an atom-dimer-scattering Efimov
resonance for positive scattering length was observed in a mixture
of atoms and halo dimers in an optically trapped gas of Caesium
atoms. Efimov resonances were also observed in a mixture of
potassium and rubidium atoms for positive and negative
scattering lengths \cite{Weber-Efimov}. However, despite all the
experimental effort, the most convincing signature of Efimov
physics, namely, equally spaced resonances in three-body observables
on the $\ln|a_0|$ scale, is yet to be seen.

There exists a large body of theoretical work on different aspects
of the Efimov physics (see \cite{Fedorov-Jensen, Hammer} and
references therein). Majority of the theoretical effort was directed
at three-body recombination processes \cite{theory}. Collisions of an
atom with a weakly bound molecule comprising identical (fermionic)
atoms were considered in Ref.\ \cite{Esry-molecule}.

In all experiments known to us \cite{Grimm,Knoop,Weber-Efimov},
Efimov states show up as resonances in the dependence of the loss
rate on the magnetic field. The goal of this paper is to point to
another possibility: thermalization rate for cold atomic mixtures
should exhibit a similar resonant behavior. There may well exist
cases when resonant losses are unobservable due to unfavorable
three-body parameters \cite{Efimov,Hammer}, and resonant
thermalization becomes a natural means of ``catching'' the Efimov
states. In this paper, we consider scattering of a heavy atom off a
molecule comprising a light and an identical heavy atom and show
that resonances due to intermediate Efimov states are equally
present in the elastic and inelastic cross sections. We focus on the
situation where the heavy atoms are bosons, and the molecules exist
in a cold atomic mixture due to an interspecies Feshbach resonance
\cite{boson-fermion,Marzok}. The atom-molecular cross sections are
calculated in the Born-Oppenheimer approximation. In connection with
the Efimov states the Born-Oppenheimer approximation was firstly
discussed in paper \cite{Fonseca}. Applying this approximation to
the scattering problem we express the cross section of three-body
collisions in terms of the scattering amplitudes corresponding to
the molecular terms (potentials) in which the three-body complex
moves. Another simplification is the use of $s$-wave scattering
approximation for the said molecular terms as well as for the
two-body interactions. Such approximation is justified for a slow
motion of the incident atom, characteristic of ultracold collisions.

\begin{figure}[b]
\includegraphics[width=0.75\columnwidth]{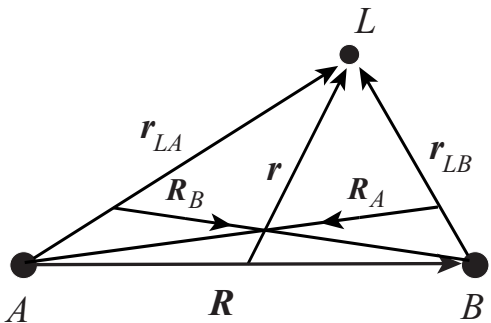}
\caption{Kinematics of the three-body interaction.}
\label{fig:Schema}
\end{figure}

\noindent{\em Statement of the problem.\/}--- We analyze a collision of a heavy atom $A$ with a molecule
$\left\{BL \right\}$ formed by one heavy atom $B$ and one light atom
$L$. Kinematics of the problem is illustrated in Fig. 1. To start with, we
disregard the fact that $A$ and $B$ are identical bosons.
We can then distinguish ``straight'' and rearrangement collisions,
\eqA{ A+\left\{BL\right\}\rightarrow A+\left\{BL\right\},\ \
A+\left\{BL\right\}\rightarrow B+\left\{AL\right\}.
}{
\nonumber 
}%
We assume that the energy of the incident atom $A$ is insufficient to break the molecule,
so that the channel where all three particles break free is closed. In a real experiment
such channel is open due to four-body collisions, but we assume that it can be neglected.

The Schr\"odinger equation for a particle $L$ with
mass $m$, interacting with two particles $A$ and $B$ of
mass $M$ reads
\eqA{
    \left[-\frac{\hbar^2}{M}\frac{\partial^2}{\partial {\bf
    R}^2}-\frac{\hbar^2}{2\mu}\frac{\partial^2}{\partial {\bf
    r}^2}+
    U_0\left({\bf r}_{LA}\right)+U_0\left({\bf r}_{LB}\right)
    \right]\Psi=E\Psi.
}{%
\eqlabel{2a} 
}%
Here, ${\bf R}$ is the distance between the heavy particles, ${\bf
r}$ is the position of the light particle relative to the center of
mass of $A$ and $B$, ${\bf r}_{LA}={\bf r}_{L}-{\bf r}_{A}={\bf
r}+{\bf R}/2$ and ${\bf r}_{LB}={\bf r}_{L}-{\bf r}_{B}={\bf r}-{\bf
R}/2$ are the positions of the light particle relative to the heavy
ones, $\mu=2Mm/(2M + m)\approx m$ is the reduced mass of the light
particle, and $E$ is the total energy of system. The choice of
coordinates is illustrated in Fig. 1.

Equation (\ref{eq:2a}) applies if $R=|{\bf R}|\ll R_0$, where
$R_0$ is the range of direct heavy-heavy interactions omitted in (\ref{eq:2a}).
The light-heavy potential $U_0$ is characterised by the two-body scattering length
$a_0$ in the zero-range approximation: $a_0\gg r_0$, where $r_0$ is the range of $U_0$.
We assume that $a_0$ is positive, i.e., that there exists a weakly bound state of the
light and heavy atoms. For overall consistency we should also  assume that (cf.\
\cite{Efimov,Hammer})
\eqA{
r_0\leq R_0 \ll a_0 .
}{%
\eqlabel{19b} 
}%

\noindent{\em The Born-Oppenheimer approximation.\/}--- In this approximation \cite{LL},
the light particle, described by the wave function $\chi({\bf r};{\bf R})$,
moves in a two-well potential, 
\eqM{
    \left[-\frac{\hbar^2}{2\mu}\frac{\partial^2}{\partial {\bf
    r}^2}+U_0\left({\bf r}+\frac{{\bf R}}{2}\right)+
    U_0\left({\bf r}-\frac{{\bf R}}{2}\right)
    \right]\chi({\bf r};{\bf R}) \\
        =\varepsilon({\bf R})\chi({\bf r};{\bf R}),
}{%
\eqlabel{3a} 
}%
where ${\bf
R}$ is regarded a
parameter. In the zero-range approximation for $U_0$ the
symmetric and antisymmetric solutions $\chi^{(\pm)}({\bf
r};{\bf R})$ read \cite{Baz,Demkov,Shlyap}
\eqM{
    \chi^{(\pm)}({\bf r};{\bf R})=\left(\frac{1}{2}\frac{1}{1\pm
e^{-\kappa_{\pm}R}}\right)^{1/2} \\ \times
    \left[\psi_{\kappa_{\pm}}(|{\bf r}-{\bf R}/2|)\pm \psi_{\kappa_{\pm}}(|{\bf r}+{\bf R}/2|)
    \right],
}{%
\eqlabel{5a} 
}%
where $\psi_{\kappa}(r)=\sqrt{\kappa/2\pi}\exp(-\kappa r)/r$, and
$\kappa_{\pm}=\kappa_{\pm}(R)$ are related to the bound state
energies as $\varepsilon^{(\pm)}=-{\hbar^2\kappa_{\pm}^2}/{2\mu} $.
Their dependence on $R$ follows from the equations
\eqA{
    \pm\,e^{-\kappa_{\pm}R}=\kappa_{\pm}R-R/a_0\; .
}{%
\eqlabel{7a} 
}%
We now look for a solution of Eq. (\ref{eq:2a}) in the
form
\eqA{
    \Psi({\bf r},{\bf R})=F^{(+)}({\bf R})\chi^{(+)}({\bf r};{\bf R})+
    F^{(-)}({\bf R})\chi^{(-)}({\bf r};{\bf R}).
}{%
\eqlabel{8a} 
}%
Substituting this Ansatz in Eq. (\ref{eq:2a}) gives rise to two independent equations for the functions
$F^{(\pm)}({\bf R})$ \cite{Mott}
\eqA{
    \left[-\frac{\hbar^2}{M}\frac{\partial^2}{\partial {\bf
    R}^2}+\varepsilon^{(\pm)}(R)\right]F^{(\pm)}({\bf R})=
    EF^{(\pm)}({\bf R}).
}{%
\eqlabel{9a} 
}%

\noindent{\em Elastic cross section.\/}--- We look for solutions of (\ref{eq:9a}) with the standard
scattering behavior for large $R$ \cite{LL,Mott}
\begin{equation}
 \label{function F-condition}
    F^{(\pm)}({\bf R})\big|_{R\rightarrow\infty}=
    e^{i{\bf k}{\bf R}}+\frac{f^{(\pm)}(\vartheta)}{R}\,e^{ikR} ,
\end{equation}
where $\vartheta$ is the angle between vectors ${\bf k}$
and ${\bf R}$. This solution corresponds to the total energy $E=
\varepsilon_0+\hbar^2k^2/M$, consisting of the binding energy of the
light particle $\varepsilon_0=-\hbar^2/2ma_0^2$, and of the energy
of relative motion of the incident heavy atom and the molecule.
Using solutions $F^{(\pm)}({\bf R})$ as building blocks, one can
construct a properly symmetrized three-body wave function for identical heavy atoms \cite{Mott}.
Of importance to us is its
asymptotic form for large $R$,
\eqM{
    \Psi({\bf R},{\bf r})=e^{-i{\bf k}{\bf R}}\psi({\bf r}_{LB})+e^{i{\bf k}{\bf R}}\psi({\bf
    r}_{LA})
    \\
    +\frac{e^{ikR}}{2R}\left[f^{(A)}(\vartheta)\psi({\bf r}_{LA})+
    f^{(B)}(\vartheta)\psi({\bf r}_{LB})\right].
}{%
\label{wave-function-symmetric-asymptotic} 
}%
where
\eqA{
f^{(A)}=f^{(+)}(\vartheta)+f^{(+)}(\pi-\vartheta)+f^{(-)}(\vartheta)-f^{(-)}(\pi-\vartheta), \\
f^{(B)}=f^{(+)}(\vartheta)+f^{(+)}(\pi-\vartheta)-f^{(-)}(\vartheta)+f^{(-)}(\pi-\vartheta).
}{%
\eqlabel{1a} 
}%
In deriving this we used approximations for the reduced masses,
$Mm/(M+m)\approx m$ and
$M(m+M)/(m+2M)\approx M/2$, valid for $m\ll M$. We also
neglected the distinction between the positions of heavy atoms
and centers of mass of the corresponding molecules, cf.\ Fig.\ \ref{fig:Schema}.
This is justified so far as the additional phase
factor in the wave function is very small, $k(m/M)r_{LB}\sim
k(m/M)r_{LA}\sim (m/M)ka_0\ll 1$. This inequality coincides with the
validity criterion of the Born-Oppenheimer approximation. The latter
is applicable if the velocity of the relative motion $v\sim \hbar
k/M$ is small compared to that of the light atom bound to the
heavy atom, $v_{L}\sim \hbar/ma_0$, i.e. $(m/M)(ka_0)\ll 1$.

To calculate the elastic cross section we note that, as  $R\to \infty $,
$\psi({\bf r}_{LB})$ and $\psi({\bf r}_{LA})$ become orthogonal
as functions of ${\bf r}$. Hence the incident flow of heavy atoms is
determined by two independent contributions both
equalling $2\hbar k/M$.
By the same reason the flow of scattered atoms is
equal to $(\hbar k/2M)[|f^{(A)}|^2+|f^{(B)}|^2]$, resulting in the
elastic cross section
\eqA{
    \sigma=\frac{\pi}{4} \int\left(|f^{(A)}(\vartheta)|^2+|f^{(B)}(\vartheta)|^2\right)\sin\vartheta d\vartheta.
}{%
\eqlabel{17b} 
}%
\noindent{\em Three-body scattering length.\/}--- A closer
inspection of \Eq{7a} shows that for $R>a_0$ both molecular terms
$\varepsilon^{(\pm)}(R)$ exponentially approach $\varepsilon _0$,
i.e., the range of the atom-molecular interaction is of the order of
$a_0$. Assuming that $k a_0\ll 1$, the $s$-wave scattering
approximation is also applicable to the atom-molecular interactions.
We are interested in a double-resonant situation, when not only the
light-heavy interaction is resonant, but also the atom-molecular
interaction becomes resonant due to an emerging Efimov state. This
limits our analysis to vicinity of scattering resonances related to
Efimov states. In this case only the $s$-wave amplitudes matter,
which are spherically symmetric. As a result $f\FM(\vartheta )$
cancels in \Eqs{1a}, and we find
\eqA{
    f^{(A)}=f^{(B)}=-\frac{2}{1/a_0^{(+)}+ik},
}{%
\eqlabel{4a} 
}%
where $a_0^{(+)}$ is the $s$-wave scattering length for the  molecular term
$\varepsilon^{(+)}(R)-\varepsilon_0$.
Under the Born-Oppenheimer approximation we have thus effectively reduced the three-body problem to a two-body one.

\noindent {\em Radial law.\/}--- It is convenient to work with a dimensionless form of (\ref{eq:9a})
for the ``plus'' term,
\eqA{
    \frac{d^2}{d\rho^2}\,u(\rho)-V(\rho)u=0.
}{%
\eqlabel{11b} 
}%
Here, $\rho=R/a_0$, $u(\rho)=F(a_0\rho)\rho$, and
\begin{equation}
 \eqlabel{14b}
    V(\rho) =
    -\frac{M}{2m}\left[\frac{G^2(\rho)}{\rho^2}-1\right],
\end{equation}
where $G(\rho)=\rho\kappa_{+}(\rho)a_0$. Eq.\ (\ref{eq:7a}) then
reads $\exp(-G)=G-\rho$. Note that \Eq{11b} is written for $k=0$, or
$E=\varepsilon_0$ \cite{Flugge}. For large $\rho $ the solution of
\Eq{11b} behaves as $u(\rho )\propto 1-\rho a_0/a_0^{(+)}$
\cite{Flugge}, so that
\begin{equation}
 \label{eq:scattering-length}
    \frac{a_0^{(+)}}{a_0} =\lim_{\rho\rightarrow\infty}
    \left[\rho-\frac{u(\rho)}{du/d\rho}\right].
\end{equation}

For $\rho\ll 1$, the potential $V(\rho)$ in \Eq{14b} behaves as
$-(s_0^2+1/4)/\rho^2$, where
\begin{equation}
 \label{s0}
    s_0=\sqrt{G^2(0)M/2m-1/4}\; ,
\end{equation}
and $G(0)\approx 0.5671$.
Consequently the general solution to (\ref{eq:11b}) for $R_0/a_0\ll\rho \ll 1$ reads,
\eqA{
u(\rho\rightarrow 0)\sim
\sqrt{\rho}\sin(s_0\ln(\Lambda_0 a_0\rho )) ,
}{%
\eqlabel{15b} 
}%
where $\Lambda_0$ is a constant. It is determined by the boundary
condition at $\rho\rightarrow R_0/a_0$, and plays the role of the
so-called three-body parameter, containing all the necessary
information about the short-range interactions. Importantly,
$\Lambda _0$ does not depend on $a_0$. Indeed, if $\rho \ll 1$,
$a_0$ in fact cancels in \Eq{11b}. This can also be seen in
\Eq{15b}: $a_0\rho =R$, and $a_0$ only occurs in the overall
coefficient. Furthermore, by allowing $\Lambda _0$ to be complex, we
can also include the information about the losses due to transitions
from the weakly bound heavy-light molecular state into deep diatomic
ones. The range $R_0$ then characterizes the ``black box'' within
which the whole of short-range physics is contained. Condition
(\ref{eq:19b}) ensures consistency of the whole viewpoint, cf.\
\cite{Efimov,Hammer}.

The ratio $a_0^{(+)}/a_0$ is a
periodic function of $\ln a_0$, which is a particular case
of the "radial law" \cite{Efimov,Hammer}. Indeed, let $u_{1,2}$ be
two linearly independent solutions of Eq. (\ref{eq:11b}), such that
$u_1(\rho)= \sqrt{\rho}\cos(s_0\ln\rho)$ and
$u_2(\rho)= \sqrt{\rho}\sin(s_0\ln\rho)$ for
$\rho\ll 1$. The solution coinciding for $\rho \ll 1$ with (\ref{eq:15b}) reads
\eqA{ u(\rho)\sim
\sin(s_0\ln\Lambda_0a_0)u_1(\rho)+\cos(s_0\ln\Lambda_0a_0)u_2(\rho).
}{
\nonumber 
}%
For $\rho \to \infty$, $u_{1,2}(\rho)=\alpha_{1,2}+\beta_{1,2}\rho$,
where the coefficients $\alpha_{1,2}$ and $\beta_{1,2}$ are determined by the potential
$V(\rho)$; they depend only
on the mass ratio $M/m$. By direct calculation with \Eq{scattering-length} we have,
\eqA{
    \frac{a_0^{(+)}}{a_0}=\alpha+\beta\cot(s_0\ln(a_0/a_*)+i\eta_*),
}{%
\eqlabel{scattering-length u1-u2} 
}%
where
$\alpha=-(\alpha_1\beta_1+\alpha_2\beta_2)/(\beta_1^2+\beta_2^2)$
and $\beta=(\alpha_1\beta_2-\alpha_2\beta_1)/(\beta_1^2+\beta_2^2)$.
Instead of one complex parameter $\Lambda_0$ we have introduced two
real parameters $a_*$ and $\eta_*$ by the equation,
$s_0\ln(\Lambda_0a_*)=-\arctan(\beta_2/\beta_1)+i\eta_*$.

\noindent{\em Results and discussion.\/}--- With losses \Eq{17b} applies to the elastic cross section $\sigma_e$,
while the inelastic $\sigma_r$ is found as a disbalance between the incoming and outgoing waves.
In the limit $k |{\rm
Im} a_0^{(+)}| \ll 1$,
$\sigma_e=4\pi|a_0^{(+)}|^2$ and $\sigma_r=(4\pi/k)|{\rm
Im}\,a_0^{(+)}|$ \cite{LL}, and we find
\eqA{ \Z \sigma_e=4\pi a_0^2(\alpha^2+\beta^2)
\frac{\sin^2(s_0\ln(a_0/a_*)+\theta_0)
+\sinh^2(\eta_*)}{\sin^2(s_0\ln(a_0/a_*))+\sinh^2(\eta_*)}, \\
\Z \sigma_r=\frac{2\pi a_0}{k}
    \frac{\beta\sinh(2\eta_*)}{\sin^2(s_0\ln(a_0/a_*))+\sinh^2(\eta_*)}\,,
}{%
\eqlabel{18b} 
}%
where $\theta_0=\arctan(\beta/\alpha)$.
The parameters $\alpha
,\beta,s_0$ and $\theta _0$ in (\ref{eq:18b}) are known functions of
the mass ratio $M/m$, while $a_*$ and $\eta_*$ are in essence
fitting parameters: $a_*$ is the value of the scattering length for
which the atom-molecular cross section has a Efimov resonance, while
$\eta_*$ determines its width. The experimentally controllable
parameter is the two-body scattering length $a_0$.

\begin{figure}[t]
\includegraphics[width=0.95\columnwidth]{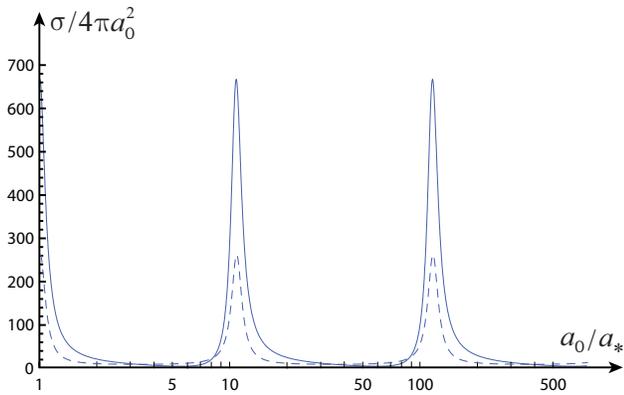}
\caption{The elastic cross section $\sigma_{e}/4\pi a_{0}^2$ (solid
line), and the inelastic one $\sigma_{r}/4\pi a_{0}^2$ (dashed line),
as functions of $a_0/a_*$ for the $^{87}{\rm Rb}$--$^{7}{\rm
Li}$ mixture, with  $ka_{0}=\eta_*=0.1$.} \label{fig:2}
\end{figure}

In Fig.\ \ref{fig:2}, $\sigma_e$ and $\sigma_r$ (\ref{eq:18b}) are
plotted as functions of $a_0/a_*$ (on a logarithmic scale) for a
mixture of $^{87}{\rm Rb}$ and $^{7}{\rm Li}$  ($M/m\approx 12.43$)
\cite{Marzok}. For this system, $s_0=1.322$, $\alpha =2.17$, $\beta
=2.55$, and $\theta_0 =0.87$. As functions of $\ln(a_0/a_*)$,
$\sigma_e$ and $\sigma_r$ are periodic with the period
$\exp(\pi/s_0)\simeq 10.8$. The graphs exhibit a typical series of
equidistant resonances. The losses are maximal at
$s_0\ln(a_0/a_*)=\pi n$, $n=0,\pm 1,\pm 2,..$, while maxima of $\sigma _e$
are somewhat shifted.

The $^{7}{\rm Li}-{}^{87}{\rm Rb}$ mixture appears to be a good
candidate for observing multiple Efimov resonances. Firstly, this
mixture has a large mass ratio, and, consequently, a relatively
small separation between Efimov resonances. Secondly, this mixture
exhibits a sufficiently wide ($\Delta B=175\,{\rm G}$) Feshbach
resonance near the magnetic field $B_0=649\,{\rm G}$ \cite{Marzok}.

While the inelastic cross section determines resonant losses, the
elastic one manifests itself through, e.g., the resonant dependence
of the thermalization rate $\gamma $ for the atom--molecular
mixture, $\gamma\propto\sigma_e$ \cite{Marzok-thermalization}. For
the $^{7}{\rm Li}-{}^{87}{\rm Rb}$ mixture, the maxima of the
elastic and inelastic cross sections are connected by the formula,
\begin{equation}
 \label{elastic and inelastic cross section ratio}
    \frac{\sigma_e^{\rm max}}{\sigma_r^{\rm max}} = 2.6\frac{ka_0}{\eta_*}.
\end{equation}
The Efimov resonances thus manifest themselves either as increased
losses or accelerated thermalization; these two ways of observing
Efimov resonances are complementary.


\noindent{\em Acknowledgements.\/}--- The authors are deeply indebted to
G. Shlyapnikov and D. Petrov for numerous enligtening discussions and
comments on the manuscript,
to C.~Marzok for a discussion of experimental techniques, and to A. Wolf
and A. Zhukov for discussions and technical assistance.
LIP and MAE are grateful
to Laboratoire de Physique Th\`eorique et Mod\`eles Statistiques,
CNRS, Universit\`e Paris Sud, for generous hospitality. WPS, MAE and MYI
acknowledge support of the Alexander von Humboldt Stiftung, BB of the
scholarship ``Mathematical Analysis
of Evolution, Information and Complexity'' at Ulm University. WPS also
acknowledges support of the Max Planck Society. This work was supported
in part by a grant
from the Ministry of Science, Research and Arts of Baden-W\"urttemberg.

\end{document}